\input harvmac.tex
%
\figno=0
\def\fig#1#2#3{
\par\begingroup\parindent=0pt\leftskip=1cm\rightskip=1cm\parindent=0pt
\baselineskip=11pt
\global\advance\figno by 1
\midinsert
\epsfxsize=#3
\centerline{\epsfbox{#2}}
\vskip 12pt
{\bf Fig. \the\figno:} #1\par
\endinsert\endgroup\par
}
\def\figlabel#1{\xdef#1{\the\figno}}
\def\encadremath#1{\vbox{\hrule\hbox{\vrule\kern8pt\vbox{\kern8pt
\hbox{$\displaystyle #1$}\kern8pt}
\kern8pt\vrule}\hrule}}

\overfullrule=0pt

%

\def\np#1#2#3{{\it Nucl. Phys.} {\bf B#1} (#2) #3}
\def\pl#1#2#3{{\it Phys. Lett. }{\bf B#1} (#2) #3}
\def\prl#1#2#3{{\it Phys. Rev. Lett.}{\bf #1} (#2) #3}

\font\zfont = cmss10 

\def\bigone{\hbox{1\kern -.23em {\rm l}}}
\def\ZZ{\hbox{\zfont Z\kern-.4emZ}}

\def\a{\alpha}

\def\e{\epsilon}

\def\m{\mu}
\def\n{\nu}

\def\r{\rho}

\def\o{\over}

\def\bbox{{\sqcap \!\!\!\! \sqcup}}

\Title{
{\vbox{
\rightline{\hbox{hepth/0107044}}
\rightline{\hbox{CALT-68-2334}}
\rightline{\hbox{UMD-PP-01-063}}
}}}
{\vbox{
\hbox{\centerline{Supersymmetry Breaking,}}
\hbox{\centerline{${\cal M}$-Theory and Fluxes}}}}
\smallskip
\centerline{Katrin Becker\footnote{$^1$}
{\tt beckerk@theory.caltech.edu}} 
\smallskip
\centerline{\it California Institute of Technology 452-48, 
Pasadena, CA 91125}

\smallskip
\centerline{Melanie Becker\footnote{$^2$}
{\tt melanieb@physics.umd.edu}}
\centerline{\it Department of Physics, University of Maryland}
\centerline{\it College Park, MD 20742-4111}

\bigskip
\bigskip
\baselineskip 18pt

\bigskip

\noindent

We consider warped compactifications of ${\cal M}$-theory to three-dimensional
Minkowski space on compact eight-manifolds. 
Taking all the leading quantum gravity corrections
of eleven-dimensional supergravity into account we obtain the solution
to the equations of motion and Bianchi identities. 
Generically these vacua are not supersymmetric
and yet have a vanishing three-dimensional cosmological constant.

\Date{July, 2001}

\newsec{Introduction}
For a long time it has been known that string theory suffers
from the vacuum selection problem
\ref\dsa{M.~Dine and N.~Seiberg, ``Couplings and Scales in
Superstring Models'', \prl {55} {1985} {366}.},
\ref\dsb{M.~Dine and N.~Seiberg, ``Is the Superstring
Weakly Coupled?, \pl {162} {1985} {299}.}.
Different shapes and sizes of the compactified dimensions lead to
many physically inequivalent degenerate vacua. This
is not a very attractive situation.
Definite predictions for all the dimensionless
constants of nature can only be made if string theory has
a unique vacuum state.
Thus over the years different mechanisms have been developed to address
this situation\foot{For a review see e.g.
\ref\dine{M.~Dine, ``Seeking the Ground State of String Theory'',
hep-th/9903212. }.} but no clear progress had been made so far.
Gukov, Vafa and Witten
\ref\gvw{S.~Gukov, C.~Vafa and E.~Witten, ``CFT's from Calabi-Yau
Four-Folds'', \np {584} {2000} {69}, hep-th/9906070.}  
realized \foot{See also
\ref\drs{K.~Dasgupta, G.~Rajesh and S.~Sethi, 
``M Theory, Orientifolds and G-Flux'', JHEP 9908:023, 1999, 
hep-th/9908088.}.}
that if we consider a warped compactification of 
${\cal M}$-theory on eight-manifolds with non-vanishing fluxes 
for tensor fields\foot{ 
Recent work on theories which include non-vanishing 
fluxes was done in \ref\gukov{S.~Gukov, ``Solitons, Superpotentials
and Calibrations'', \np {574} {2000} {169}, hep-th/9911011.},  
\ref\silver{E.~Silverstein, ``(A)DS Backgrounds from
Asymmetric Orientifolds'', hep-th/0106209. },
\ref\lust{G. Curio, A. Klemm, B. Kors and D. L\" ust, 
``Fluxes in Heterotic and Type II String Compactifications'', 
hep-th/0106155.},
\ref\grana{M. Grana and J. Polchinski, ``Gauge/Gravity 
Duals with Holomorphic Dilaton'', hep-th/0106014.},
\ref\bertolini{M. Bertolini, V. L. Campos, G. Ferretti, P. Fre,
P.~Salomonson and M.~Trigiante,  
``Supersymmetric Three Branes on Smooth ALE Manifolds with Flux'', 
hep-th/0106186. } and
\ref\warner{T. Eguchi, N. Warner and S-K. Yang, 
``ADE Singularities and Coset Models'', hep-th/0105194.}.}
 the expectation values for the complex structure 
and K\"ahler structure moduli fields are no longer arbitrary.
Most of them are fixed in terms of the discrete fluxes found in
\ref\Becker{K.~Becker and M.~Becker,``${\cal M}$-Theory on
Eight Manifolds'', \np {477} {1996} {155}, hep-th/9605053. }
and
\ref\svw{S.~Sethi, C.~Vafa and E.~Witten, ``Constraints on Low Dimensional
String Compactifications'',
\np {480} {1996} {213}, hep-th/9606122.}.
As shown by Giddings, Kachru and Polchinski
\ref\gkp{S.~B.~Giddings, S.~Kachru and J.~Polchinski,
``Hierarchies from Fluxes in String Compactifications'',
hep-th/0105097.} a similar situation appears in  
the Type IIB theory.

In this paper we will be interested in finding all vacua
for warped compactifications
of ${\cal M}$-theory on compact eight-manifolds.
Compact manifolds are of special interest as they lead to a finite
three-dimensional Planck scale.
Taking all the leading quantum gravity corrections
of ${\cal M}$-theory into account it is our goal to derive the most general
solution to the equations of motion for compactifications 
on eight-dimensional K\"ahler manifolds to 
three-dimensional Minkowski space.
This solution will be written in terms of first order constraints
which will be much easier to solve than the second order constraints
coming from the equations of motion. We hope this will be useful
in order to construct new interesting concrete models in the future.
Generically we will find solutions which have a vanishing three-dimensional
cosmological constant and broken supersymmetry. Such an interesting
situation has appeared recently in the no-scale models of
{\gkp}.

In section two we will derive the
solution to the equations of motion and Bianchi identities
for compactifications of ${\cal M}$-theory to
three-dimensional Minkowski space.
In section three we will summarize our solution.
In section four we discuss the constraints imposed by
supersymmetry and the possibility to break supersymmetry to
$N=0$ by turning on some
specific fluxes. In section five we will review
the interpretation of the 
flux constraints that our solution obeys and the relation
to the moduli space problem of ${\cal M}$-theory compactifications.
Many of the moduli fields can be stabilized once the constraints
are taken into account.
We will finish in section six with some concluding remarks. 
\newsec{Solution to the Equations of Motion}

The bosonic part of the action of eleven-dimensional supergravity
\ref\cjs{E.~J.~Cremmer, B.~Julia and J.~Scherk, 
``Supergravity Theory in Eleven Dimensions'', 
\pl {5} {1978} {409}.}
including the leading quantum corrections  
\ref\greena{M.~B.~Green and P.~Vanhove, 
``D Instantons, Strings and M Theory'', \pl {408} {1997} {122},
hep-th/9704145.},
\ref\greenb{M.~Green, M.~Gutperle and P.~Vanhove, 
``One Loop in Eleven Dimensions'', 
\pl {409} {1997} {177}, hep-th/9706175.},
\ref\dlm{M.~J.~Duff, J.~T.~Liu and R.~Minasian, 
``Eleven Dimensional Origin of String/String Duality: A One 
Loop Test'', \np {452} {1995} {261}, hep-th/9506126.}, 
\ref\tseytlin{ A.
A. Tseytlin, ``$R^4$ Terms in 11 Dimensions and Conformal Anomaly
of (2,0) Theory'', \np {584} {2000} {233}, hep-th/0005072. }
has the following form 
\eqn\ai{ \eqalign{ & S = S_0+S_1 ,\cr & S_0=-{1 \over 2
\kappa^2}\int d^{11}x \sqrt{-g} \left[R -{1\over 2 \cdot 4!}
F^2-{1\over 6 \cdot 3! \cdot (4!)^2}\e_{11}{ C} F F
\right],\cr & S_1=-b_1 T_2 \int d^{11} x \sqrt{-g} (J_0 -{1\over
2} E_8) +T_2 \int C \wedge X_8. \cr } }
Here $b_1={1\over {(2 \pi)^4 3^2 2^{13}}}$ and $T_2$ is the
membrane tension related to the eleven-dimensional Newton's
constant by
\eqn\aai{
T_2=\left( { 2\pi^2 \over \kappa^2 }\right)^{1/3}.
}
We will be using the conventions of {\tseytlin}.
Furthermore, $F=dC$ is the four-form field strength and $J_0$,
$E_8$ and $X_8$ are quartic polynomials in the eleven-dimensional
Riemann tensor. The explicit form of the polynomial $J_0$ is
\eqn\ci{ J_0  =3\cdot 2^8 (R^{HMNK}R_{PMNQ}{R_H}^{RSP}{R^Q}_{RSK}+
{1\over 2} R^{HKMN}R_{PQMN}{R_H}^{RSP}{R^Q}_{RSK})+O(R_{MN}). }
The polynomial $E_8$ is an eleven-dimensional generalization of the 
eight-dimensional Euler integrant and is given by
\eqn\cii{
E_8  ={ 1\over 3!} \e^{ABCM_1 N_1 \dots M_4 N_4}
\e_{ABCM_1' N_1' \dots M_4' N_4' }{R^{M_1'N_1'}}_{M_1 N_1} \dots
{R^{M_4' N_4'}}_{M_4 N_4}.
}
Capital letters range over $0,\dots,10$. The expression for $X_8$
is
\eqn\ciii{
X_8={ 1\over 192 (2 \pi)^4} \left[ {\rm tr} R^4 -{1 \over 4}
({\rm tr}R^2)^2 \right] .
}
The Einstein equation which follows from this action is
\eqn\aii{ R_{MN}-{1\over 2}g_{MN} R -{1 \over 12} T_{MN}= -\beta {1\over \sqrt{-g}}
         {\delta \over \delta g^{MN}} \left(
\sqrt{-g} (J_0-{1\over 2}E_8) \right), }
where $T_{MN}$ is the energy momentum tensor of $F$ given by
\eqn\aiii{
T_{MN}=F_{MPQR} {F_N}^{PQR}-{1 \over 8} g_{MN}F_{PQRS}^2, 
}
and we have set $\beta=2 \kappa^2 b_1 T_2$. 

Without sources the field strength obeys the Bianchi identity
\eqn\aiiixx{ dF=0, }
and the equation of motion
\eqn\aiiixxi{ d*F={1\over 2}F \wedge F+ {{\beta} \over {b_1}}X_8.}

In the following we shall be interested in considering
compactifications on eight-manifolds.
Our goal is to derive the general conditions under which the
equations of motion have a solution by a perturbation expansion
in $t$, where $t$ is the radius of the eight-manifold which is taken 
to be large .
Such a large radius expansion was used in
\ref\witten{E.~Witten and L.~Witten,``Large
Radius Expansion of Superstring Compactifications '', \np {281}
{1987} {109}. } for compactifications of the heterotic string.
We consider the background metric to be a warped
product {\Becker}
\eqn\aiv{
ds^2= e^{2A(y)}
\eta_{\mu \nu }dx^{\mu}dx^{\nu} + e^{2B(y)} g_{mn}dy^m dy^n ,
}
where $\eta_{\m\n}$ describes the three-dimensional Minkowski space $M_3$.
The metric $g_{mn}$ is taken to be of order $t^2$
\eqn\avxx{g_{mn}=t^2{g_{mn}^{(0)}}+ {g_{mn}^{(1)}}  \dots, }
and describes the eight-dimensional internal manifold $Y_4$.
In our notation the indices $m,n,\dots$ are real.
In this paper
we will be interested in compactifications where $Y_4$ is 
K\"ahler. It would be interesting to find the generalization of
our analysis to non-K\"ahler manifolds such as the $Spin(7)$ holonomy
manifolds considered in
\ref\katrin{K.~Becker, ``A Note on Compactifications on $Spin(7)$
Holonomy Manifolds'', JHEP 0105:003, 2001, hep-th/0011114.}.
To derive the three-dimensional equations of motion by a 
perturbative expansion we need to analyze the scaling behaviors 
of all fields as function of the radius.
From {\avxx} it follows that
the inverse metric scales as $g^{mn} \sim 1/t^2$, the
Riemann tensor scales with $t^2$
and the scalar curvature  is of order  $t^{-2}$
\eqn\avxxix{R=g^{mn}R_{mn}=g^{mn}g^{kl}R_{mknl}=
t^{-2}R^{(0)}+t^{-8}R^{(1)}+
\dots.}
Furthermore, the Ricci tensor is of order zero
\eqn\avxxz{R_{mn}={R_{mn}^{(0)}}+ t^{-6}{R_{mn}^{(1)}}+  \dots, }
while from {\cii} we find that the quartic 
polynomial of the Riemann tensor
scales as 
\eqn\aviv{E_8(Y_4)=t^{-8}{E_8}^{(0)}(Y_4)+\dots,}
and a similar expansion for $J_0$.
To leading order in the large $t$-expansion one can
replace the Riemann tensor appearing in $J_0$ {\ci} by the Weyl tensor.
This will be useful later on.

In compactifications with
maximally symmetric three-dimensional space-time the field
strength is a sum
\eqn\aivxi{ F=F_1+F_2, }
where $F_1$ has the form
\eqn\aaiv{F_{\mu\nu\rho m}=\e_{\m\n\r} \partial_m f, }
with indices on the three-dimensional Minkowski space while $F_2$
has only indices on the eight-manifold. Here $f=f(y)$ is a function
of the internal coordinates that will be determined later on.
This form of the field strength follows from
Poincar\'e invariance. The above ansatz for
$F_1$ satisfies the Bianchi identity for the external
component of the tensor field.  

In order to derive the field equations order by order
in the $t$-expansion we will make the following ansatz
for the scaling behavior
of the tensor fields
\eqn\aaaivx{
f=f^{(0)}+t^{-6}f^{(1)}+\dots
}
and
\eqn\aaivx{
F_2={F_2}^{(0)}+t^{-6}{F_2}^{(1)}+\dots. }
From {\aaiv} and {\aaaivx} we see that $F_1$ has a similar
expansion as $F_2$.
Also, we will be making an ansatz for the 
scaling behavior of the warp factors
\eqn\aviv{e^{X}=1+{X^{(1)}\over t^6}+\dots,}
with $X=A,B$.

Combining the
leading orders of the external and internal Einstein's equations
we see that the internal manifold is Ricci flat
\eqn\axixzp{R^{(0)}_{mn}=0.}
Also, the external component of the flux vanishes to leading order
because $f^{(0)}=const$.

To order $t^{-8}$ the external component of Einstein's equation is
\eqn\ax{ -4\bbox A^{(1)}-14\bbox B^{(1)}-
{1\over 48} ({F_2}^{(0)})^2+R^{(1)}+{\beta \o 2} {E_8}^{(0)}(Y_4) =0. }
Here we used the fact that we can
neglect the contribution of the warp factor to the $(\rm
Riemann)^4$ terms.
Thus the right hand side
of {\aii} can be evaluated on a
product space of the form $M_3\times Y_4$. To obtain
the contribution coming from $E_8$ we have taken into account that
for these product spaces we have\ref\louis{M.~Haack and J.~Louis, ``M-Theory
Compactified on Calabi-Yau Fourfolds with Background Flux'',
hep-th/0103068,}
\eqn\aaxi{ E_8(M_3 \times Y_4) =-E_8(Y_4) -8 R(M_3) E_6(Y_4). }
Here $E_6(Y_4)$ is the cubic polynomial of the internal
Riemann tensor
\eqn\axixxzixz{E_6(Y_4)=
2^8(R_{a \;\; c}^{\;\; b \;\; d} R_{d \;\; b}^{\;\; e \;\; g}
R_{e \;\; g}^{\;\; a \;\; c}+
R_{a \;\; b}^{\;\; c \;\; d} R_{d \;\; e}^{\;\; b \;\; g}
R_{g \;\; c}^{\;\; e \;\; a}).}
At this point
we will be assuming that the internal manifold is K\"ahler so that 
we can introduce complex coordinates which we will be denoting by
$a, b, {\bar a}, {\bar b},\dots$. 
Since $R(M_3)$ is the scalar curvature of the external space
the second term in the previous equation actually vanishes.
To evaluate the contribution
from $J_0$ to the external Einstein equation we have used the fact that
$J_0$ is the sum of an external and an internal part. The
external part vanishes because the Weyl tensor vanishes
identically in three dimensions \ref\misner{C.~W.~Misner,
K.~S.~Thorne and J.~A.~Wheeler, ``Gravitation'', W.H.Freeman and
Company, New York (1973).}.
The internal part does not contribute because it vanishes for Ricci
flat K\"
ahler manifolds.
This can be easily checked using the explicit expression for
$J_0$ appearing in {\ci}.

We now would like to consider the
order $t^{-6}$ of the internal Einstein equation. Let us start
with the $(a,\bar b)$ 
component which takes the form 
\eqn\axixz{ { {R_{a{\bar b}}^{(1)}}-{1\over 2}{g_{a{\bar b}}^{(0)}}R^{(1)}
-3{\partial}_a{\partial}_{\bar b}}C^{(1)}+3{g_{a{\bar
b}}^{(0)}}{\bbox}C^{(1)}
-{1\over 12}{T_{a{\bar b}}^{(1)}}
=\beta {\partial}_a{\partial}_{\bar b}
E_6(Y_4). }
Here we have introduced the notation $C^{(1)}=A^{(1)}+2B^{(1)}$.
To evaluate the right hand side of {\aii} we used the identity
\eqn\axiixzx{ {\delta\over \delta g^{a{\bar b}}} J_0=-
\partial_a \partial_{\bar b} E_6(Y_4),}
which is valid for Ricci flat K\"ahler manifolds. 
This
can be checked using the results of
\ref\gz{M.~T.~Grisaru and D.~Zanon,
``Sigma-Model Superstring Corrections to the
Einstein Hilbert Action'', \pl {177} {1986} 347.}, 
\ref\fpss{M.~D.~Freedman, C.~N.~Pope, M.~F.~Sohnius and K.~S.~Stelle,
``Higher-Order $\sigma$-Model Counterterms and the Effective Action
for Superstrings'', \pl {178} {1986} 199.} and
\ref\gvz{M.~T.~Grisaru, A.~E.~M.~van den Ven and D.~Zanon,
``Four-Loop Divergences for the $N=1$ Supersymmetric Non-Linear
Sigma-Model in Two Dimensions'', \np {277} {1986} 409.}
or by a
lengthy but straigthforward calculation.
There is one point with which one has to be careful though,
which is the scheme dependence of $J_0$.
The explicit form of the terms that involve the Ricci tensor in
{\ci} can be changed using the equations of motion.
This issue has been discussed in detail in the literature 
for the Type IIA higher order interactions. 
We have done the above calculation in the same scheme that was used in
{\gz}, {\fpss} and {\gvz} or more concretely in 
\ref\sen{A.~Sen, ``Central Charge of the Virasoro Algebra 
for Supersymmetric Sigma Models on Calabi-Yau Manifolds'', 
\pl {178} {1986} {370}.}.

Taking the trace of {\axixz} with the metric 
$g^{(0)}_{a\bar b}$ we obtain an expression for the
scalar curvature of the internal manifold
\eqn\azzx{R^{(1)}=7{\bbox}A^{(1)}+14{\bbox}B^{(1)}-
{\beta \o 3}{\bbox}E_6(Y_4).}
Here we have used that the energy-momentum tensor is 
traceless in eight dimensions.
Inserting this into the external Einstein equation {\ax} we obtain a
determining equation for the warp factor $A^{(1)}$
\eqn\axiz{ 3\bbox A^{(1)}-{1\over 48} ({F_2}^{(0)})^2
-{\beta \over 3}\bbox E_6(Y_4)+
{\beta \o 2} {E_8}^{(0)}(Y_4) =0. }

The $F_1$ equation of motion states
\eqn\anm{ \bbox f-{1\over 48}F_2^{(0)}{\tilde \star}F_2^{(0)}
+{\beta \o 2}
E_8^{(0)}(Y_4)=0,}
where by ${\tilde \star}$ we mean the Hodge dual with respect to the
eight-dimensional metric.
Substracting this from  equation {\axiz} and integrating
over the compact eight-manifold we obtain the condition that $F_2^{(0)}$
has to be self-dual
\eqn\axiixzxiz{F_2^{(0)}=\tilde \star F_2^{(0)}.}

Since $F_2^{(0)}$ is self-dual we can compare {\axiz} with
{\anm} to get a relation between the external
component of the tensor field, the warp factor $A^{(1)}$ and the
polynomial $E_6$
\eqn\axn{f^{(1)}=3A^{(1)}-{\beta \o 3} E_6(Y_4)+const.}

There is an integrability condition for being able to solve equations
{\axiz} and
{\anm} for $A^{(1)}$ and $f$ respectively {\witten}.
The source terms must be orthogonal to the
zero modes of the operator $\bbox$. The only zero modes of the operator
$\bbox$ on a compact manifold are constants, so that the integrability
condition for both equations becomes
\eqn\aaxiiizx{ {\int}_{Y_4}  F_2^{(0)}\wedge F_2^{(0)} +{\chi \over 12}=0,}
where $\chi$ is the Euler number of the eight-manifold.
This condition has been found before in
{\Becker} and
{\svw}. It indicates that compactifications on eight-manifolds
with non-vanishing Euler number are only consistent if fluxes
are turned on.

Having shown the self-duality of $F_2^{(0)}$
let us go back to the internal Einstein equation {\axixz}.
It turns out that
any self-dual tensor in eight dimensions
satisfies
\ref\DW{B.~De Wit, ``Properties of SO(8) Extended Supergravity'',
\np {158} {1979} {189}.}
\eqn\axii{ F_{mpqr}{F_n}^{pqr}={1 \over 8} g_{mn} F_{pqrs}^2. }
Due to this identity the energy momentum tensor $T_{mn}^{(1)}$ vanishes
identically, so that it does not contribute to the internal
Einstein's equations. The equation {\axixz} then becomes 
\eqn\zxy{{R_{a{\bar b}}^{(1)}}-{1\over 2}
{g_{a{\bar b}}^{(0)}}{\bbox}( C^{(1)}-{\beta \o 3} E_6(Y_4))=
3{\partial_a \partial_{\bar b}}
( C^{(1)}+{\beta \o 3} E_6(Y_4)).}
Recall that for a K\"ahler manifold the Ricci tensor and
the metric are curl free. Taking the curl of {\zxy}
gives 
\eqn\zpx{{\partial}_a \bbox ( C^{(1)}-{\beta \o 3} E_6(Y_4))=0 .}
For a compact eight-manifold the solution to this equation
is
\eqn\axixxz{ 2B^{(1)}+A^{(1)}-{\beta \over 3} E_6(Y_4)=const, }
This determines the warp factor
$B^{(1)}$ in terms of $A^{(1)}$.

Furthermore we observe that to this order
the internal manifold is no longer Ricci flat because
the Ricci tensor satisfies 
\eqn\axi{ { {R_{a{\bar b}}^{(1)}}=2\beta 
{\partial}_a{\partial}_{\bar b}}
E_6(Y_4). }
This fact is familiar from the Type IIA theory in which the 
background metric is no longer Ricci flat to the next to leading
order in the ${\a}'$ expansion once higher order interactions
are taken into account {\sen}. 
This completes our discussion of the $(a,{\bar b})$ component of the
internal Einstein equation.

The remaining Einstein equation takes the form
\eqn\yyy{{R_{ab}^{(1)}}-3{\partial}_a{\partial}_{b}C^{(1)}
 +\beta  {\partial}_a{\partial}_{ b} 
E_6(Y_4)=0,}
and a similar expression for the antiholomorphic component.
Here we have taken into account 
\eqn\axiixzx{ {\delta\over \delta g^{ab}} J_0=
\partial_a \partial_{ b} E_6(Y_4),}
and the same result holds for the variation 
with respect 
to a metric with two antiholomorphic indices.
With the 
solution {\axixxz} for $C^{(1)}$ these equations become
\eqn\csd{  {R_{ab}^{(1)}}={R_{{\bar a}{\bar  b}}^{(1)}}=0,}
as it has to be for the metric to be K\"ahler.

It was shown in 
\ref\ns{D.~Nemeschansky and A.~Sen, 
``Conformal Invariance of Supersymmetric ${\sigma}$-Models on
Calabi-Yau Manifolds'', \pl {178} {1986} {365}.},
\ref\cfpss{P.~Candelas, M.~D.~Freeman, C.~N.~Pope, M.~F.~Sohnius
and K.~S.~Stelle, ``Higher Order Corrections to Supersymmetry
and Compactifications of the Heterotic String'', 
\pl {177} {1986} {341}.}
and {\sen} that there always exists a K\"ahler metric on a
Calabi-Yau manifold which satifies Einstein's equations
{\axi} and {\csd} even if the manifold is no 
longer Ricci flat.  We will be assuming that 
$Y_4$ is a Calabi-Yau manifold so that supersymmetry is not broken 
by the background metric but by the fluxes. 
It would be
interesting to know if non-K\"ahler manifolds solve the 
next to leading order constraints.

Finally, the equation of motion for the
internal component of the tensor field  ${F_2}^{(0)}$ is
\eqn\axiv{ d(\tilde \star {F_2}^{(0)}) = 0. }
Since ${F_2}^{(0)}$ is closed and self-dual
this equation is always satisfied and imposes no further
constraints.

\newsec{Summary of the Solution to the Equations of Motion}

The solution to the
equations of motion and Bianchi identity 
for ${\cal M}$-theory compactified to three
dimensional Minkowki space on an eight-dimensional
K\"ahler manifold is characterized by the following
conditions.

\item{$\triangleright$} The field strength is of the form 
\eqn\nn{F=F_1+F_2,}
where $F_1$ is the external component
given by {\aaiv} and $F_2$ has only indices on the internal 
eight-manifold. 
\item{$\triangleright$ } To leading order the internal component 
of the field strength must be self-dual
\eqn\axivxz{\tilde \star {F_2}^{(0)} = {F_2}^{(0)}.}
and satisfy the integrability condition
\eqn\nmp{ {\int}_{Y_4}  F_2^{(0)}\wedge F_2^{(0)} +{\chi \over 12}=0,}
where $\chi$ is the Euler number of the eight-manifold.
\item{$\triangleright$}The leading order  
the external component of the field strength vanishes 
\eqn\nmp{{F_1}^{(0)}=0,}
while the next to leading order component ${F_1}^{(1)}$
is related
to the warp
factor $A^{(1)}$ by equation {\axn}.
\item{$\triangleright$} 
The warp factors $A^{(1)}$ and $B^{(1)}$ follow from equations
{\axiz} and {\axixxz}.
\item{$\triangleright$} To leading order the internal manifold $Y_4$
is Ricci flat.
To the next to leading order the internal manifold 
is no longer Ricci flat. The Ricci tensor is given by
{\axi} and {\csd}. These equations have a solution if 
$Y_4$ is a Calabi-Yau manifold.

Let us analyze the conditions under which the internal
component of the field strength $F_2$ is self-dual
{\louis}. The
behavior under duality of a four-form on an eight-dimensional
K\"ahler manifold is the following
\eqn\avi{
\tilde \star f_{(4,0)}=f_{(4,0)}\qquad
\tilde \star f_{(3,1)}=-f_{(3,1)}\qquad
\tilde \star f_{(1,3)}=-f_{(1,3)}\qquad
\tilde \star f_{(0,4)}=f_{(0,4)},
}
where in general $f_{(p,q)}$ denotes a form of type $(p,q)$ with $p$
holomorphic and $q$ antiholomorphic indices.
In order to derive this result it is easiest to use the following
representation of the epsilon tensor
\eqn\aavi{
\e_{a\bar b c\bar d e \bar f g \bar h }=
g_{a \bar b } g_{c \bar d} g_{e \bar f } g_{g \bar h }\pm
{\rm permutations}.
}
From {\avi} it follows that the self-duality 
constraint {\axivxz} imposes the conditions
\eqn\nmz{F_{(1,3)}=F_{(3,1)}=0.}

However, the constraint allows a $(2,2)$ form
\eqn\nmzi{F_{(2,2)}=f_{(2,2)},}
which is primitive
\eqn\mmx{J \wedge f_{(2,2)}=0,}
where $J$ is the K\"ahler form 
of the manifold to leading order. This is so 
because every primitive $(2,2)$ form is self-dual. Of course, 
$f_{(2,2)}$ should be closed in order for the Bianchi identity
to be satisfied. Notice that for an eight-manifold a self-dual
$(2,2)$ form is not necessarily primitive. This situation
is rather different than for threefolds where for $(2,1)$ forms 
primitivity and
self-duality are equivalent.
For a fourfold a self-dual $(2,2)$ form does not have to be primitive but a
primitive $(2,2)$ form is self-dual.
Finally, the constraint {\axivxz}
 allows a $(2,2)$ form which is not primitive 
\ref\gh{P.~Griffiths and
J.~Harris, ``Principles of Algebraic Geometry'', Wiley, New York
(1978).}
\eqn\avii{ F_{(2,2)}=J \wedge J f_{(0,0)}, }
with $f_{(0,0)}$ closed.

Altogether, the equations of motion and Bianchi identities
will be satisfied for ${F_2}^{(0)}$ of the form
\eqn\aix{ {F_2}^{(0)}=f_{(4,0)}+f_{(0,4)}
+f_{(2,2)}+J \wedge J
f_{(0,0)}. }
This summarizes all the conditions describing the solution
to the equations of motion and Bianchi identities. We now would
like to compare with the constraints coming from supersymmetry. 

\newsec{Supersymmetry and Supersymmetry Breaking}

The solution that we just presented does not need to
be supersymmetric. Let us recall the constraints imposed
by supersymmetry on these compactifications. In {\Becker}
it was shown that for a supersymmetric compactification
of ${\cal M}$-theory on eight-manifolds the four-form is of type $(2,2)$,
i.e. 
\eqn\axvi{
F_{(4,0)}=F_{(0,4)}=F_{(3,1)}=F_{(1,3)}=0.
}
Further the 
non-vanishing component of $F$ has to be primitive
\eqn\mmp{
F_{(2,2)} \wedge J=0.
}
Therefore, supersymmetry only allows a four-form flux that takes the form
\eqn\azm { {F_2}^{(0)}=
f_{(2,2)}. }
Comparing with the result coming from
the equations of motion {\aix} we see that there is
the interesting possibility 
that supersymmetry can be broken by turning on the 
$(4,0)$ form (or the corresponding $(0,4)$ form)
\eqn\axvip{
f_{(4,0)}\neq 0,
}
or a $(2,2)$ form that is not primitive.
From {\aix} we see that such a non-primitive $(2,2)$ form is
\eqn\pxp{ F_{(2,2)}= J
\wedge J f_{(0,0)}. }

In both cases we know from the results of this paper that
even if supersymmetry is broken after turning on these fluxes
the three-dimensional cosmological constant vanishes.
Such an interesting scenario was first discussed in
the context of supersymmetry breaking by gluino
condensation in the heterotic string in
\ref\drsw{M.~Dine, R.~Rohm, N.~Seiberg and E.~Witten,
``Gluino Condensation in Superstring Models'',
\pl {156} {1985} {55}.}. Supersymmetry is broken in these
models by giving an expectation value to the holomorphic three-form
of the Calabi-Yau threefold.
More recently Giddings, Kachru and Polchinki 
{\gkp} found the realization
of this scenario in the context of ${\cal F}$-theory compactifications.
In fact, the models described in {\gkp} can be obtained
from our models by a specific choice of eight-manifold that
is an elliptic fibration over a threefold.  

Let us mention briefly some concrete examples of compactifications 
of ${\cal M}$-theory  on eight-manifolds that have appeared in the 
literature. 
All these examples involve non-compact internal manifolds and the 
relevant part of the ${\cal M}$-theory action {\ai} is $S_0$. 
A supersymmetric model in which
$F=f_{(2,2)}$ where $f_{(2,2)}$ is primitive
can be obtained by taking the internal space to be 
the eight-dimensional Stenzel metric
\ref\Cvetic{M.~Cvetic, G.~W.~Gibbons, H.Lu and C.~N.~Pope,
``Ricci-Flat Metrics, Harmonic Forms and Brane Resolutions'',
hep-th/0012011.}.
A solution which breaks supersymmetry because
the four-form is not primitive is given by the self-dual harmonic form
on the complex line bundle over $CP^3$. It would certainly 
be interesting if concrete examples involving compact internal
manifolds could be constructed since these models give rise 
to a finite three-dimensional Newton's constant.
Some supersymmetric examples along these lines were constructed
in {\drs}. 

\newsec{Flux Constraints and Stabilization of Moduli Fields}
Since the work of Dine and Seiberg 
it is well known that in string theory it is difficult to
stabilize the moduli fields {\dsa}, {\dsb}.
Different shapes and sizes of the compactified dimensions 
lead to many physically inequivalent degenerate vacua.
Recently Gukov, Vafa and Witten {\gvw} found an interesting 
interpretation of the supersymmetry constraints {\axvi} and {\mmp}.
It was observed that the constraint 
\eqn\axvi{
F_{(4,0)}=F_{(0,4)}=F_{(3,1)}=F_{(1,3)}=0,
}
can be used to stabilize
the complex structure moduli fields. This is because 
given a flux which satisfies the Dirac quantization 
condition the complex
structure of the eight-manifold has to be adjusted in such a way that
the constraint equations are satisfied.
Furthermore, the condition
\eqn\mmp{
F_{(2,2)} \wedge J=0,
}
can be used to fix many of the 
K\"ahler moduli of the internal manifold once the flux
$F_{(2,2)}$ is used as an input. The radius of the eight-manifold
(which is a K\"ahler modulus) cannot be determined though. The reason for
this is that the equations are invariant under a rescaling with this
parameter. 
 
A corresponding interpretation of the constraints for 
Type IIB compactifications on six-manifolds {\drs}
was found in {\gkp}.
Here a very nice derivation was made in terms of the supergravity
potential. From this calculation it becomes clear how the discrete 
fluxes determine most of the moduli fields even if the vacua are not
supersymmetric. 

It would be interesting to derive the constraints found
in this paper from a supergravity potential along the lines of 
{\gkp}. The corresponding potential has been
computed in {\louis}.

\newsec{Conclusion}
In this paper we have found warped compactifications of ${\cal M}$-theory 
on (compact) eight-manifolds which generically are not supersymmetric
and yet have a vanishing three-dimensional cosmological constant.
Our calculation was based on a perturbative expansion in terms
of the radius of the eight-manifold and took all the leading
quantum corrections of ${\cal M}$-theory into account.
Many of the moduli fields appearing in these compactifications can
be stabilized using the constraints on the fluxes. These 
constraints have to be 
obeyed for the equations of motion and Bianchi identity
to be satisfied.

It would certainly be interesting to extend the analysis 
performed in this paper to the next order 
in perturbation theory. It is conceivable that the compactification
radius can be fixed in this way.

In order to do this calculation one first has to 
determine additional terms in the effective action 
of ${\cal M}$-theory. 
So for example to compute the next to leading order of
the equation of motion for $F_2$ an additional
term in the ${\cal M}$-theory action {\ai} becomes relevant
\eqn\bi{ 
S_3\propto \int \sqrt{-g}F^2 R^3. 
}
This term has been considered previously in the literature \tseytlin, 
\ref\strom{A. Strominger, ``Loop Corrections to the Universal
Hypermultiplet'',  \pl {421} {1998} {139}, hep-th/9706195 . } but 
the coefficient of this interaction has not been 
determined so far. 

However, it is also possible that non-perturbative
                                                                                                   effects of the form considered in
\ref\bbs{K.~Becker, M.~Becker and A.~Strominger,
``Five-Branes, Membranes and Nonperturbative String Theory'',
\np {456} {1995} {130}, hep-th/9507158.}
and 
\ref\wsp{E.~Witten, ``Non-Perturbative Superpotentials in
String Theory'', \np {474} {1996} {343}, hep-th/9604030.} 
stabilize the radius.

\vskip 2cm
\noindent {\bf Acknowledgement}

\noindent 
We would like to thank especially E.~Witten for many useful
discussions and suggestions. We also would like to thank 
S.~Giddings, J.~Polchinski, J.~Schwarz and A.~Tseytlin for useful 
discussions. 
The work of K.~Becker was supported by the U.S. Department of Energy 
under grant DE-FG03-92-ER40701. M.~Becker is supported in part by 
the Alfred Sloan Foundation.

\listrefs

\end